\documentclass[useAMS,usenatbib]{mn2e}
 
\usepackage{epsf}

\def\eps@scaling{.95}
\def\epsscale#1{\gdef\eps@scaling{#1}}
\def\plotone#1{\centering \leavevmode
    \epsfxsize=\eps@scaling\columnwidth \epsfbox{#1}}

\def\plotfiddle#1#2#3#4#5#6#7{\centering \leavevmode
    \vbox to#2{\rule{0pt}{#2}}
    \includegraphics{#1}}

\newcommand{\Msolar}{M$_{\odot}$}

\newcommand{\kms}{\,km s$^{-1}$}
\newcommand{\mic}{\,$\mu$m}

\newcommand{\brg}{Br$\gamma$}
\newcommand{\htwo}{H$_2$\,2.122\,$\mu$m}
\newcommand{\fe}{[FeII]\,1.644\,$\mu$m}
\newcommand{\irasa}{IRAS\,17150-3224}
\newcommand{\irasb}{IRAS\,17423-1755}
\newcommand{\irasc}{IRAS\,17441-2411}
\newcommand{\irasd}{IRAS\,21282+5050}

\title[Near-Infrared Echelle Spectroscopy of PPN]
{Near-Infrared echelle spectroscopy of Proto-Planetary Nebulae: 
probing the fast wind in H$_2$}

\author[C.J. Davis et al.]
         {
  C.J. Davis$^{1}$, M.D. Smith$^{2}$, T.M. Gledhill$^{3}$ \& 
  W.P. Varricatt$^{1}$ \\
  $^1$Joint Astronomy Centre, 660 North A'oh\={o}k\={u} Place, 
	University Park, Hilo, Hawaii 96720, USA. \\
  $^2$Armagh Observatory, College Hill, Armagh BT61 9DG, 
             Northern Ireland. \\
  $^3$Department of Physics, Astronomy \& Mathematics, University of Hertfordshire, 
    Hatfield, Herts AL10 9AB, U.K. 
	  }

\date{Accepted ...
      Received ...;
      in original form ...}
\pagerange{\pageref{firstpage}--\pageref{lastpage}}
\pubyear{2005}

\begin{document}
\maketitle
\label{firstpage}

\begin{abstract}

Echelle spectroscopy of \htwo , \fe\ and \brg\ line emission from a
very young planetary nebula, \irasd , and from four proto-planetary
nebulae, IRAS~19343+2926 (M\,1-92), \irasa\ (AFGL~6815), \irasb\
(Hen~3-1475), and \irasc\ is presented.  H$_2$ line
emission is detected in discrete shock fronts in the lobes of each
nebula, regardless of source spectral type (although non-detections in
IRAS~09371+1212 (Frosty Leo) support claims that late spectral types
do not produce bright H$_2$ line emission).  In \irasa\ we also uncover
possible signs of rotation, as would be expected if the H$_2$ features
were excited in a magneto-centrifugal disk wind.  \fe\ emission was
detected in only one source, M\,1-92 (notably, the source with the
brightest H$_2$ features).  Again, the emission is predominantly
excited in high-velocity shocks in the bipolar lobes of the PPN.
The H$_2$ and [FeII] observations of M\,1-92, and
the complex H$_2$ profiles in \irasd , are explained using the shock
models of Smith and collaborators. We show that bow shock models are
generally able to account for the observed line profiles, peak
velocities, the double-peaked profiles in \irasd\ and the spatial
distribution of H$_2$ and [FeII] in M~1-92.  J-type bow models are adequate 
in each case, i.e. a strongly-magnetised wind is not required. 

Finally, \brg\ is detected in each of the five targets; in absorption
in the G-type PPN though in emission in the O and B-type sources.
\brg\ emission is detected predominantly toward the near-IR continuum
peak in each PPN, with only very weak emission detected in the
extended lobes of the O- and B-type sources. In \brg , low peak
velocities, though very broad profile widths, are measured in each
target, regardless of nebula inclination angle.  The emission must
therefore derive from ionised regions in a fast wind very close to the
central star (rather than from shocks in the bipolar lobes), or, in
the late-type sources, from absorption in an equatorial torus.

\end{abstract}


%

\section{Introduction}

The proto-planetary nebula (PPN) phase is a short-lived episode in the
rapid evolution of stars from the asymptotic giant branch (AGB) to the
planetary nebula (PN) phase (Kwok 1993).  The central stars of PPN are
still too cool to ionise the slow-moving circumstellar shell ejected
during the AGB phase, yet they appear to drive collimated,
high-velocity winds which shock and shape this shell, and almost
certainly entrain slow-moving AGB ejecta to produce a fast molecular
wind (Lee \& Sahai 2003).

From recent observations, particularly high-resolution imaging studies
(e.g. Sahai et al. 1998; Bujarrabal et al. 1998a; Sahai, Bujarrabal \&
Zilstra 1999; Ueta, Meixner \& Bobrowski 2000; Su, Hrivnak \& Kwok
2001), it has become clear that the rapidly-evolving PPN phase
ultimately shapes the morphology of the resulting PN.  At some point
during or soon after the detachment of the AGB envelope, the mass-loss
geometry changes from more-or-less spherically symmetric to axially
symmetric.  Bipolar morphologies, knotty jets and Herbig-Haro
(HH)-like ``bow shocks'' are all manifestations of this evolutionary
process. Indeed, they seem to be present even in relatively ``young''
PPN.

Recently, Bujarrabal et al. (2001) have found that radiatively-driven
winds cannot account for the high momentum and energy implied by their
CO observations of PPN fast winds.  This has prompted theorists to
look beyond the classical ``interacting stellar winds'' model of Kwok
et al. (1978) and investigate whether an accretion disk scenario,
similar to that used to explain jets from young stars and active
galactic nuclei, could be used to explain both the high degree of
collimation and the point symmetry seen in many PPN jets (Soker \&
Rappaport 2000; Frank \& Blackmann 2004).  In these models, the
accretion disk must form through binary interactions.
Magneto-centrifugal launching from the surface of the disk is then a
means of converting gravitational energy into kinetic energy in a fast
wind.  If magneto-hydrodynamics (MHD) do dictate the energetics and
collimation of PPN outflows (see also Garc\'{\i}a-Segura et al. 1999;
Blackman et al. 2001), then they will also define the physics of the
shocks in these flows, an effect which can be investigated with
high-resolution observations of the emission regions associated with
the shocks.

We therefore observed a modest number of PPN and young PN at high
spectral resolution.  Our sample includes reddened PPN (\irasa ,
\irasc , IRAS~09371+1212 [Frosty Leo] and IRAS~17106-3046) as well as
more evolved PPN and young PN (M~1-92,
\irasb\ and \irasd ), sources with earlier spectral types that are on
the brink of the transition to the PN stage. We sought observations in
a number of atomic and molecular gas tracers.  H$_2$ 1-0S(1) emission
at 2.122\mic\ is an important probe of excited molecular gas in the
detached AGB envelope ($>$90\% of the AGB ejecta will be
atomic/molecular), while \brg\ recombination line emission is
potentially a useful tracer of the fast, ionised wind. [FeII] at
1.64\mic\ could also be an excellent tracer of moderate-to-high
excitation shocks in the fast wind and/or the interaction zone between
the wind and AGB envelope (the Fe+ ionisation fraction peaks at
$T\sim14,000$~K for $n_H \sim 10^5$cm$^{-3}$; [FeII] has been shown to
be a useful diagnostic in studies of HH shocks,
e.g. Nisini et al. 2002; Davis et al. 2003a).  In the dense,
inner-core regions of each PPN, these near-IR lines are also less
affected by extinction than their optical counterparts, and so are
expected to be ideal probes of the dynamical interaction between the
fast wind and AGB envelope.

\begin{table*}
        \caption{UKIRT/CGS4 echelle spectroscopy observing log}
        \begin{tabular}{llcccccccccc}
        \hline  
IRAS   & Other & Central$^1$ & $D^2$ &  RA$^3$  & Dec$^3$  & Date$^4$ & Date$^4$ & Date$^4$ & Slit$^5$ & $\theta^6$ & Ref$^7$ \\
number & name  & star        & (kpc) & (2000.0) & (2000.0) &  (H$_2$) &  (\brg)  & ([FeII]) &   P.A.   &            &         \\

        \noalign{\smallskip}
        \hline
        \noalign{\smallskip}

21282+5050 &            & O9       & 3   &  21 29 58.5 &  51 04 00 & 040728        & 040728 & 040728 & 160\degr &      0\degr & 1 \\ 
19343+2926 & M~1-92     & B0.5V    & 2.5 &  19 36 18.9 &  29 32 51 & 040430,040705 & 040530 & 040530 & 133\degr &     35\degr & 2 \\
17423-1755 & Hen~3-1475 & Be       & 5   &  17 45 14.2 & -17 56 47 & 040615,040705 & 040624 & 040624 & 135\degr &     50\degr & 3 \\  
17441-2411 & AFGL~5385  & G0       & 4   &  17 47 13.5 & -24 12 51 & 040625,040703 & 040625 & 040705 & 197\degr &  $<$30\degr & 4 \\  
17150-3224 & AFGL~6815  & G2I      & 2.4 &  17 18 19.8 & -32 27 21 & 040615,040705 & 040625 & 040625 & 125\degr &  $<$30\degr & 5 \\  
09371+1212 & Frosty Leo & K7II     & 3   &  09 39 53.6 &  11 58 54 & 040428        &   --   &   --   & 156\degr &     15\degr & 6 \\ 
17106-3046 &            &  ?       & ?   &  17 13 51.6 & -30 49 40 & 040625        &   --   &   --   & 128\degr &     52\degr & 7  \\

        \noalign{\smallskip}
        \hline
        \noalign{\smallskip}
        \end{tabular}

  \begin{list}{}{}
  \item[$^1$] Optical spectral type of the central object 
  \item[$^2$] Estimated distance to the source
  \item[$^3$] Central source position.
  \item[$^4$] UT date of the \htwo , \brg\ and \fe\ observations 
   of each target (year/month/day).  The ``left'' and ``right'' H$_2$ slits referred
   to in Section 3 were, in most cases, observed on the second date in column 7.
  \item[$^5$] Position angle, measured East of North, of the echelle slit. 
  The same angle was used at each wavelength, and for each of the three 
  adjacent H$_2$ slit positions
  (the angle corresponds to the axis of the bipolar lobes in each PPN).
  \item[$^6$] Inclination angle of the bipolar nebula with respect to the
  plane of the sky. 
  \item[$^7$] References to $D$ and $\theta$: 
                     1) Meixner et al. 1997;
                     2) Bujarrabal et al. 1997, Bujarrabal et al. 1998b;
 	             3) Riera et al. 1995, Borkowski \& Harrington 2001;
                     4) Su et al. 1998, Weintraub et al. 1998; 
                     5) Loup et al. 1993;
                     6) Mauron et al. 1989, Roddier et al. 1995;
                     7) Kwok et al. 2000 
                           
  \end{list}
  \label{tab1}
\end{table*}

%
 
\section{Observations and Data Reduction} 
 
Echelle spectra at 1.6440\mic, 2.1218\mic\ and 2.1662\mic\ were
obtained at the 3.8\,m United Kingdom Infrared Telescope (UKIRT), on
Mauna Kea, Hawaii, on various dates in Spring 2004 (see Table 1).  The
cooled grating spectrometer CGS\,4 (Mountain et al. 1990) was used.  This
instrument is equipped with a 256$\times$256 pixel InSb array and a 31
lines/mm echelle grating.  The pixel scale is 0.41\arcsec
$\times$0.88\arcsec\ (0.41\arcsec\ in the dispersion direction).  A
2-pixel-wide slit was used, resulting in a velocity resolution of
$\sim 16$\,km s$^{-1}$ (this being the Full Width at Half Maximum
(FWHM) of unresolved sky or arc lamp lines).  

At the above wavelengths the 
 H$_2$ 1-0S(1) 
 ($\lambda_{\rm vac} =$ 2.1218334\mic ; Bragg, Smith \& Brault 1982), 
 H{\sc i} Br$\gamma$ 
 ($\lambda_{\rm vac} =$ 2.166167\mic ) and 
 [FeII] $a^4D_{7/2} - a^4F_{9/2}$ 
 ($\lambda_{\rm vac} =$ 1.643998\mic ; Johansson 1978) 
transitions are well centred on the array.  The wavelength coverage at 
each of these wavelengths was approximately
 2.115-2.128\mic, 2.158-2.173\mic\ and 1.638-1.649\mic\ 
(these ranges shifted slightly from night-to-night, depending on the
setting of the echelle grating angle).

For each target, and for each line, an exposure at a ``blank sky''
position was followed by five on-source ``object'' exposures, the
former being subtracted from all of the object frames.  This sequence
of six frames was repeated two or three times in each line to build up
signal-to-noise.  To avoid saturation on the bright continuum
associated with each PPN, relatively short exposures were used;
60\,sec ($\times$2 coadds) for the two K-band lines, and 80\,sec
($\times$2 coadds) in the H-band. The total integration time on each
source was: 30~mins in H$_2$ -- central slit (in all seven targets);
20~mins in H$_2$ -- left and right slits (in the five targets with
H$_2$ detections); 26.5~mins in [FeII] (40~mins for \irasa\ and \irasb ), 
and 30~mins in \brg .

With the above exposure times, at the high spectral resolution of the
instrument ($R \sim 18,750$), there is essentially no detectable
thermal emission (or associated shot noise) from the sky or background
between the well-resolved OH sky lines.  Thus, the process of ``sky
subtraction'' mainly serves to remove the OH lines and any ``warm''
pixels that remain after bad-pixel masking and flat-fielding (all data
frames are bad-pixel-masked, using a mask taken at the start of each
night, and flat-fielded, using an observation of an internal blackbody
lamp obtained before each set of target observations).  At 2.122\mic\
and 2.166\mic\ the H$_2$ 1-0S(1) and Br$\gamma$ lines are well
separated from any bright OH sky lines.  Thus, any residual sky lines
in the reduced spectral images resulting from imperfect sky
subtraction do not affect the data.  However, the
\fe\ line is close to an OH line, the 1.644216\mic\ 5-3~R2 transition,
so residual emission from this OH line was removed using the {\em
STARLINK} package {\em FIGARO}, specifically the {\em polysky}
routine.  This routine fits a polynomial to regions above and below
the target spectrum (along columns in the spatial/slit
direction): the fit to these sky regions is then used to subtract sky
emission, column-by-column, across the spectral image.

Coadded H$_2$ and [FeII] spectral images were wavelength calibrated
using OH sky lines, there being four lines well spaced across the
array in both cases (Rousselot et al. 2000).  The first raw object
frame observed at each wavelength on each source was used as the
reference frame.  The {\em STARLINK:FIGARO} software packages used to
do this also correct for distortion along the columns in each image
(i.e. along arc or sky lines), via a polynomial fit to the OH lines in
each row.  Examination of the OH lines in distortion-corrected and
wavelength-calibrated raw frames showed that the velocity calibration
along the length of the slit, measured from Gaussian fits to the sky
lines, is extremely good for the H$_2$ and [FeII] data.  The relative
velocity calibration along the slit and between adjacent slits and
targets is estimated to be accurate to $\sim 3$km s$^{-1}$ .

Unfortunately, only one bright OH line was evident in the raw
Br$\gamma$ spectral images, at 2.1711170\mic . A pair of weaker lines
appears in some the data at the blue end of the spectrum, and a second
pair of weak lines occurs near the centre of the array.  However,
these lines were generally too faint for accurate calibration, and
were in any case unresolved by Rousselot et al. (2000).  We therefore
used telluric absorption features in the standard star spectrum to
wavelength calibrate the central region in each \brg\ spectral image.
Note that in this case no correction was made for optical distortions
along the slit axis.  Fortunately, the \brg\ emission was confined to
a region very close to the source continuum in each spectral image, so
distortion was not considered to be a major source of uncertainty in
the wavelength/velocity calibration.  Three or four telluric lines
were observed in each standard star spectrum; these were identified in
the Arcturus atlas of Hinkle, Wallace \& Livingstone (1995) and their
wavelengths measured from the solar atlas available from the {\em BAse
de donnees Solaire Sol 2000} solar archive hosted by l'Observatoire de
Paris (http://bass2000.obspm.fr/solar{\_}spect.php).  \brg\ absorption
lines, seen in each standard, were also used (after correction for the
radial velocity of the standard). The OH lines noted above were
subsequently used to check the calibration.  The telluric features
were relatively weak in most cases, and their wavelengths only
measured to an accuracy of $\sim 0.1$ Angstrom. The velocity
calibration in the \brg\ data is therefore only considered to be
accurate to $\sim 5-10$~km s$^{-1}$.

At all three wavelengths velocities were calibrated with respect to
the {\em kinematic} Local Standard of Rest [LSR].  They were not
corrected for the systemic velocity in each region.

Finally, spectra of A or G-type standard stars were obtained and
reduced in a similar fashion.  In each case an extracted,
wavelength-calibrated spectrum was ``grown'' into a 2-D image so that
division by this image would correct each reduced PPN spectral image
for telluric absorption (\brg\ absorption lines were removed from the
standard star data before they were used to calibrate the 2.166\mic\
images).  Although most of the data were obtained in photometric
conditions, variable cirrus on 15 June, 24 June and 3 July introduced
0.5-1.5 magnitudes of attenuation.  The absolute flux calibration of
these observations (i.e. the centre H$_2$ slit in \irasa\ and \irasb , the
[FeII] and \brg\ observations of \irasb , and the left and right H$_2$
slits in \irasc ) will therefore only be accurate to within a factor of
2-3.  The seeing was in the range 0.5\arcsec --0.8\arcsec , i.e. equal to or
somewhat less than the slit width and offsets used for the adjacent
H$_2$ observations.

%
%
\begin{figure} 
\begin{center}
  \framebox{
 \epsfxsize=7.0cm
 \epsfbox{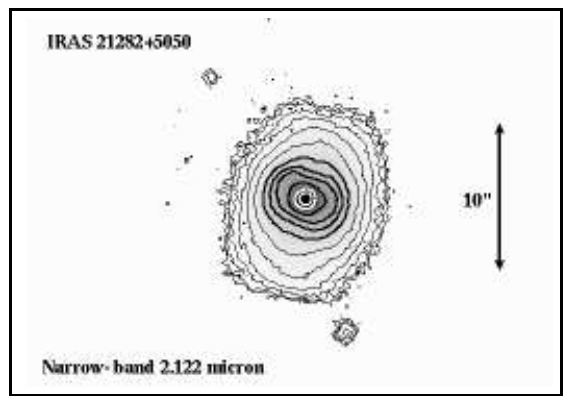}
   }
\caption{Narrow-band 2.122\mic\ image of \irasd , acquired to
establish the orientation of the bipolar lobes in this young PN.  Note that
the image is not continuum-subtracted; nor is it flux calibrated, so
the contours are arbitrary (although they approximately represent
1,2,4,8,16,32,64,96,128,192,320 and 480-times the standard deviation
in the blank-sky around the source). North is up and east is to the left. }
\end{center}
\end{figure} 
%

%

\section{Results}

%
\begin{figure*} 
\epsfxsize=15.0cm
\epsfbox[15 210 600 580]{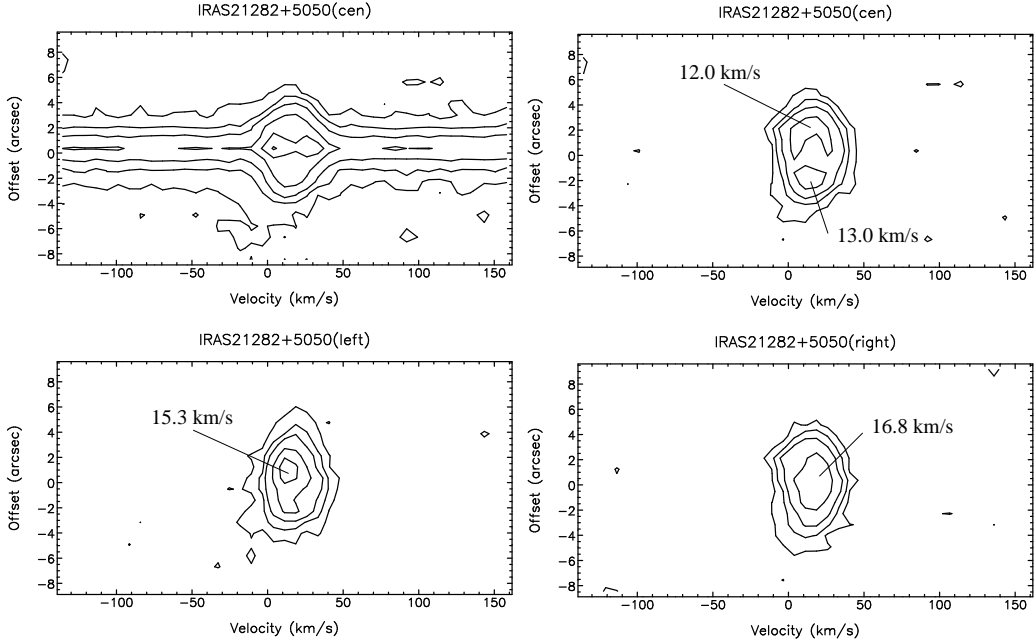}
\caption{PV diagrams showing the H$_2$ 2.122\mic\ emission in  \irasd . 
 The top of each spectral image is orientated roughly to the north
 (p.a. $\sim -20\degr$). With respect to the centre slit (cen), the
 ``left'' and ``right'' spectra were offset 1.2\arcsec\ to the east and west,
respectively. The data from the central slit are shown before and
after continuum-subtraction (top two panels); only continuum-subtracted PV diagrams are
shown for the left and right slits (bottom two panels).  The contour levels measure
0.005,0.015,0.03,0.06,0.09,0.12,0.24,0.48 Jy per arcsec along
the slit in all four plots. The radial velocities of emission peaks,
measured from Gaussian fits to spectra extracted over 3 rows (2.7\arcsec ) centred
on the peak, are labeled.  The velocity zero point is {\em not} 
corrrected for the systemic LSR velocity. }
\label{21282con}
\end{figure*} 
%

In this paper we have targeted a very young PN and six PPN; the
sources are listed in order of spectral type in Table 1.  The spectral
energy distribution (SED) of each source is double-peaked, though in
most cases dominated by a strong infrared component, as expected for a
detached, dusty shell surrounding a stellar photosphere (e.g. Kwok et
al. 1996; Ueta et al. 2000).  Bipolar lobes are evident in published,
high-resolution images of each target (except for \irasd , though see
Figure 1). The lobes in each case extend over at least a few
arcseconds in projection on the sky. In the optical, the aspect ratio
of the bipolar lobes (length/width) is typically 2--5.

For each source the spectrometer slit was centred on the bright near-IR
continuum peak; the slit was aligned along the axis through the
bipolar lobes, as defined by high-resolution images (e.g. 
M\,1-92 -- Bujarrabal et al. 1998a; 
\irasb\ -- Reira et al. 2003;
\irasc\ -- Su et al. 1998;
\irasa\ -- Kwok et al. 1998; 
IRAS~09371+1212 -- Sahai et al. 2000; 
IRAS~17106-3046 -- Kwok, Hrivnak \& Su 2000).  
We refer to this position as the ``central slit'' (cen)
position in each case. We found no high-resolution narrow-band 
images of \irasd\
in the literature (although see the K-band image of Latter et al. 1995).  
We therefore obtained a narrow-band ($\lambda /
\delta\lambda \sim 100$) 2.122\mic\ image of this target under good
seeing conditions ($\sim 0.6$\arcsec ) with the {\em UKIRT Fast Track
Imager (UFTI)} (Roche et al. 2003). An adjacent continuum image was
not obtained; nor were the data flux-calibrated.  In this image
(Figure 1), \irasd\ was seen to be slightly extended in a
north-north-west to south-south-east direction, as seems to be the
case in the K-band image of Latter et al. (1995): a position angle
(p.a.) of 160\degr\ was thus adopted for this target.

Although H$_2$ observations were obtained for seven targets, line
emission was detected in only the first five sources in Table 1.  The
upper limit to the line flux in the lobes of the two non-detections
was $\sim 5$~mJy arcsec$^{-1}$ ($\sim 1 \times
10^{-18}$~W~m$^{-2}$~arcsec$^{-2}$). When H$_2$ was detected in the
central slit position, i.e. along the polar axis of a source, H$_2$
data at adjacent, parallel slit positions were also obtained.  In each
case the slit was offset $\sim 1$\arcsec\ either side of the central
axis.  The exact offset -- reported in the H$_2$ position-velocity
(PV) diagram Figure captions -- is measured perpendicular to the slit
axis.  We refer to these offset positions as ``left'' and ``right''.
In this way, H$_2$ spectra were obtained at three adjacent, parallel
slits across the width of the first five sources listed in Table 1.

\fe\ and \brg\ observations were also then obtained for these five
targets, with the slit centred on the same continuum peak position and
using the same position angle.  In [FeII] and Br$\gamma$ only this one
central slit position was observed in each source.  \brg\ was detected
in all five targets.  However, [FeII] was only detected in M~1-92.
The upper limit to the [FeII] flux in the lobes of the \irasa , \irasb
, \irasc\ and \irasd\ nebulae was $\sim 3-10$~mJy arcsec$^{-1}$ ($\sim
0.5-2.0 \times 10^{-18}$~W~m$^{-2}$~arcsec$^{-2}$).

Below we discuss the sources individually. 
H$_2$ PV diagrams are shown for each target.  For
the central slit position a PV diagram showing the line and continuum
emission is presented along-side a continuum-subtracted PV diagram.  In the
latter the continuum has been removed from the spectral image,
row-by-row, by fitting the continuum on either side of the line
emission region with a second-order polynomial.  The PV diagram
showing both the line emission and the continuum serves to illustrate
the relative strength of the two emission components.  For the left
and right slit positions, only continuum-subtracted PV plots are
presented. In all PV diagrams positive offsets are approximately to
the north or north-west, depending on the slit position angle.  Slit
angles are listed in Table 1 and in the Figure captions.


\subsection{IRAS\,21282+5050}    

\irasd\ is probably slightly more evolved than the other targets
discussed in this paper.  It is considered to be a very young PN with
a WR central star (reclassified as O9 by Crowther, De Marco
\& Barlow 1998) surrounded by a dusty core with a central hole (Meixner et al. 
1997; 1998).  It exhibits weak radio continuum emission at 2~cm and 6~cm
(Likkel et al. 1994) and has probably evolved from a carbon-rich,
intermediate mass star.

Our narrow-band, 2.12~$\mu$m image of \irasd\ (Figure 1) shows the
target to be elongated in two directions.  The outer, faint contours
extend toward the north-north-west and south-south-east, along an
axis orientated at a position angle of about 160\degr.  We assume this
to be the orientation of the bipolar lobes and adopt this angle for
our spectroscopy.  However, the brighter contours in the core of
\irasd\ are orientated perpendicular to this axis, at a position angle
of 75\degr ($\pm$10\degr).  The near-IR image in Figure 1 bears a
striking resemblance to the 12.5\mic\ image of Meixner et al. (1997).
They observe the same elongation toward the north-north-west and
south-south-east (p.a. $\sim$165\degr ) and also, in their deconvolved
image, resolve the inner core into two peaks, aligned along an axis
orthogonal to the fainter emission (p.a. $\sim$70\degr ).  They
interpret their images and the source's spectral energy distribution
in terms of dust emission from a disk viewed edge-on, implying that
the bipolar lobes must lie close to the plane of the sky.

Hora et al. (1999) included \irasd\ in their extensive spectroscopic
survey of PN.  They obtained spectra at two positions, toward the core
and toward a position offset 3\arcsec N, 3\arcsec W.  They detected
H$_2$ emission lines from the latter.  Davis et al (2003b) later
presented an integrated spectrum of this target; they too detected
line emission from H$_2$.

%
%
\begin{figure} 
   \plotfiddle{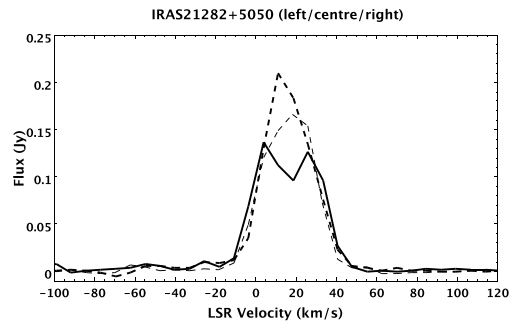}{55mm}{-90}{40}{40}{-160}{210}
\vspace*{-1.2cm}
\caption{H$_2$ spectra across the width of \irasd . The spectra
represent three rows ($\sim$2.7\arcsec ), centred on the source
continuum, extracted from the ``left'', ``centre'' and ``right'' slit
positions (plotted with thick-dash, full, and thin-dash lines,
respectively).}
\label{21282spec}
\end{figure}
%

In the central slit data in Figure \ref{21282con} we see \htwo\
emission along the slit axis and toward the source continuum position.
As we demonstrate in subsequent sections, this is somewhat unexpected,
particularly for a source with bipolar lobes orientated close to the
plane of the sky: in the other targets discussed later, the H$_2$ is
{\em confined} to discrete knots in the bipolar lobes. The H$_2$ near
the continuum position in \irasd\ is therefore unusual.  It may be
excited either very close to the central star, perhaps in the extended
wings of bow shocks in the bipolar lobes, or in a second outflow
component orientated along the line of sight and therefore in the
equatorial plane.  Equatorial H$_2$ features have been observed in at
least one other PPN, namely CRL\,2688, where multiple bow-shaped shock
structures are spatially resolved in both the bipolar lobes and in the
equatorial regions (Sahai et al. 1998).

Our three near-IR slits were orientated perpendicular to the dust ring
found by Meixner et al. (1997).  Consequently, the spatial variations
along each slit probably correspond to offsets along the bipolar lobes, and the
emission above and below the continuum in each PV diagram must
be associated with the bipolar lobes.  In other words, each
PV plot {\em could} include H$_2$ line emission from an equatorial flow 
{\em and} from the bipolar lobes.

%
\begin{figure*} 

\epsfxsize=15.0cm
\epsfbox[15 210 600 580]{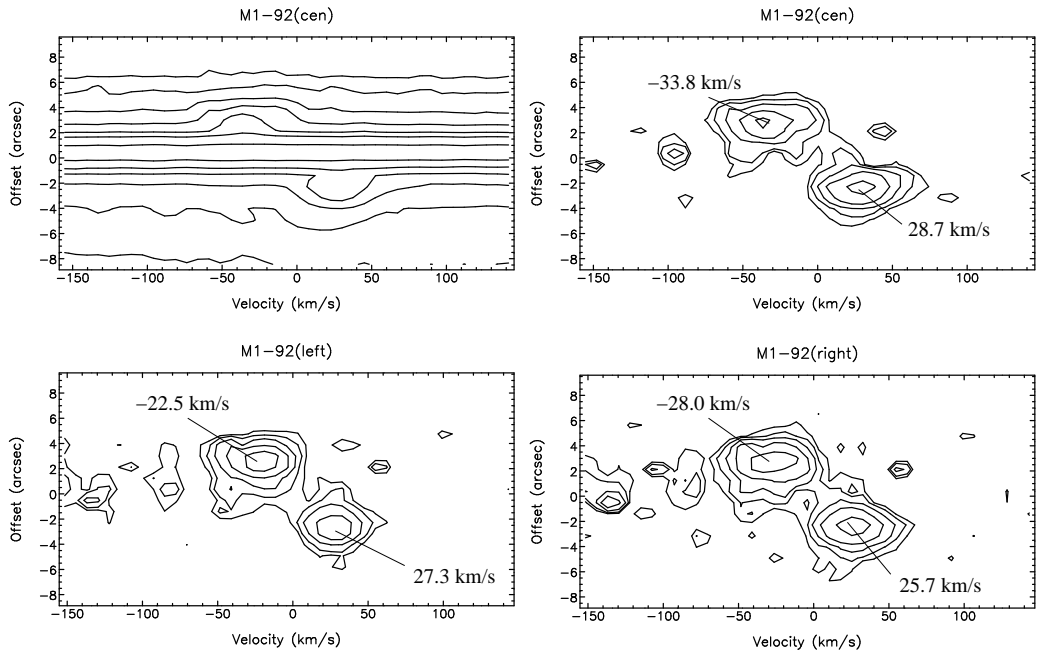}

\caption{PV diagrams showing the H$_2$ 2.122\mic\ emission in M1-92
(see Figure 2 for details). The top of each spectral
image is orientated roughly to the north-west (p.a. $\sim -47\degr$);
the ``left'' and ``right''
spectra were offset 1.2\arcsec\ to the north-east and south-west,
respectively. The contour levels measure
0.04,0.08,0.16,0.32,0.64,1.28,2.56 Jy per arcsec along
the slit in all four plots. }
\label{m192con}
\end{figure*} 
%

In the continuum-subtracted H$_2$ PV diagram for our central slit, the
emission describes a ring in position-velocity space.  From a slice
taken along the slit axis (i.e. along a single 0.52~A/7.4~\kms -wide
column in the PV diagram), the diameter of this ring is found to be of
the order of 4.4\arcsec ($\pm$0.9\arcsec ).  This diameter probably
represents the distance between emission features in the bipolar
lobes, above and below the source continuum position.  (It does not
correspond to the diameter of the dust torus observed by Meixner et
al. 1997.)

Continuum-subtracted PV diagrams from the ``left'' and ``right'' slits
are also shown in Figure~\ref{21282con}.  Spectra extracted from the
middle three rows (offset = 0\arcsec) in each of the spectral images
are shown in Figure~\ref{21282spec}.  The broadest profile is seen in
the centre slit data. Here the emission line is double peaked, and can
be well fitted with a two-component Gaussian profile with peaks
separated by 22\kms .  The spectra from the left and right slit
positions, although slightly skewed to blue and red-shifted
velocities, peak at an intermediate velocity of $\sim 16$\kms .  These
spectra -- all obtained toward the central continuum position -- may
therefore be understood in terms of: (1) emission excited in the wings of
extended bow shocks seen side-on (the double-peaked profiles
representing emission from the near and far sides of the bow), or (2)
an expanding shell or curved shock directed toward the observer.
These two scenarios will be investigated in Sect. 4.3.1, where we
discuss whether an equatorial wind is needed {\em in addition to} an
orthogonal, bipolar flow to explain the H$_2$
profiles in Figures \ref{21282con} and \ref{21282spec}.


\subsection{M\,1-92} 
 
M\,1-92 (IRAS~19343+2926) is a bipolar, oxygen-rich PPN (spectral type
B0.5V).  The relatively-high temperature of its central star ($\sim
20,000$\,K) suggests that it is probably more evolved than the
other PPN targets in Table 1.  Bujarrabal, Alcolea \& Neri (1998b) estimate
a kinematic age of about 900\,yrs for M\,1-92. The nebula comprises
two bipolar lobes of scattered, polarised light (Herbig 1975; Trammell
\& Goodrich 1996; Bujarrabal et al. 1998a).  Spectra from these lobes
consist of a stellar continuum, scattered permitted Balmer and HeI
lines, and a variety of low-excitation forbidden emission lines, the
latter being confined to fast (200-300\kms ) shocks in the bipolar
lobes (Solf 1994).  The two lobes are separated by a dark lane,
although the central star is just visible because of the inclination
of the north-western, blue-shifted lobe toward the observer.  In
optical {\em HST} images the continuum in the lobes appears smooth,
while the shock-excited [SII] and [OI] is shown to delineate knotty,
collimated jets that are roughly, though not precisely, aligned with
the axis of the nebula (Trammell \& Goodrich 1996; Bujarrabal et
al. 1998a).

Two hollow molecular shells surround the bipolar lobes (Bujarrabal et
al. 1997). Observed in CO, these probably represent the ejecta from
the AGB phase; gas swept up by the fast, post-AGB wind.  They contain
most of the nebula material and are expanding at high speed (up to
70\kms ).  Bujarrabal et al. also report the detection of fast
molecular clumps at the tips of the bipolar lobes/shells (i.e. along
the jet axes), as well as a dense, expanding, equatorial torus,
with a diameter of about 2\arcsec -3\arcsec .
 
In near-IR images of M\,1-92, knots of \htwo\ emission are observed
along the axes of the two bipolar lobes (Bujarrabal et al. 1998a;
Davis et al.  2003b). The emission from these regions is evident in
our echelle data; spectroscopy at the three parallel slit positions
across the width of M\,1-92 are shown in
Figure~\ref{m192con}. Discrete H$_2$ emission features are seen in all
three slits; H$_2$ is not detected toward the continuum position.  The
emission knot to the north-west is blue-shifted (radial velocity,
$V_{\rm LSR} \sim -30$~\kms ) while the emission in the counter-lobe
is red-shifted ($V_{\rm LSR} \sim +30$~\kms ).  When corrected for
inclination, these values convert to a velocity along the bipolar axis
of the system of about $50$~\kms , a speed that is comparable to the
velocity of the molecular AGB ejecta (Bujarrabal et al.
1998b).

%
\begin{figure} 

\epsfxsize=7.5cm
\epsfbox[85 150 550 650] {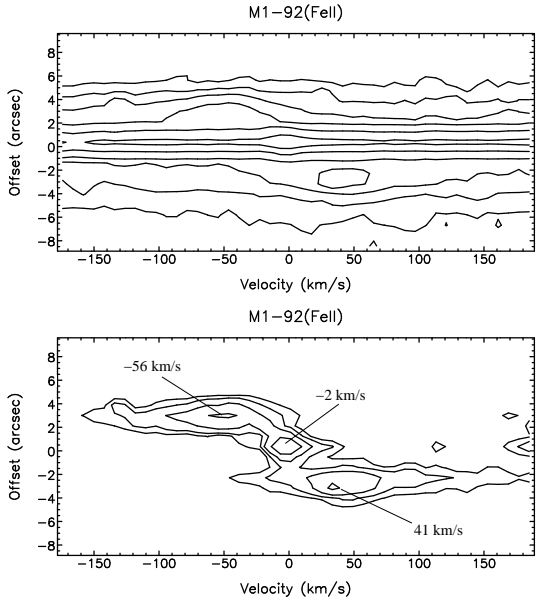}

\caption{PV diagrams showing the [FeII] 1.644\mic\ 
emission through the centre of M1-92.  
In the lower plot the continuum emission has been fitted and removed.
The LSR velocity scale has been calibrated with respect to the
wavelength of the [FeII] line.  The contour levels measure
0.005,0.01,0.02,0.04,0.08,0.16,0.32 Jy per arcsec along the slit in
both plots.}
\label{m192con2}
\end{figure} 

%
%
\begin{figure} 

    \plotfiddle{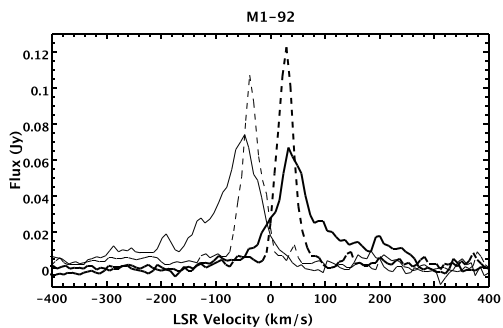}{55mm}{-90}{40}{40}{-160}{210}

\vspace*{-0.8cm}
\caption{Plots comparing H$_2$ and [FeII] line profiles above and
below the stellar continuum in Figure~\ref{m192con2}, averaged over
2.7\arcsec\ regions along the slit in both cases.  The H$_2$ data are
plotted with dashed lines, the [FeII] with solid lines. The H$_2$
spectra, extracted from the centre slit, have also been multiplied by
a factor of 0.1.}
\label{m192h2fe-spec}
\end{figure}
%

Of the five sources detected in \htwo\ emission, M\,1-92 was the only
target where \fe\ was detected.  In these [FeII] data
(Figure~\ref{m192con2}) we see blue-shifted and red-shifted features
in the bipolar lobes similar to those seen in H$_2$.  Extracted H$_2$
and [FeII] spectra are compared in Figure~\ref{m192h2fe-spec}.  These
data were extracted over the same regions in the two bipolar lobes.
The [FeII] spectra peak at slightly more blue-shifted/red-shifted
velocities ($V_{\rm LSR} \sim -56$\kms\ and $\sim 41$\kms); the [FeII]
profiles are also considerably broader.  Gaussian fits to the
extracted H$_2$ profiles yield FWHM velocities of $\sim$ 35 -- 45~\kms
; the [FeII] spectra are roughly twice as wide, and they exhibit
strong blue- and red-shifted line wings on the blue- and red-shifted
peaks, respectively.  At the 10\% peak intensity level, the [FeII]
lines are $\sim$ 330 -- 340~\kms\ wide.  

The blue and red [FeII] peaks are also slightly further apart in
separation along the slit, as compared to their H$_2$ counterparts.  From
profiles taken along the slit axis from the continuum-subtracted
spectral images, integrated over the same velocity range in each case,
the separation between the [FeII] peaks was found to be 5.53\arcsec\
compared to 5.20\arcsec\ for the H$_2$ peaks.  An error of $\sim
0.04$\arcsec\ is assumed, based on the accuracy with which we were
able to calibrate the pixel scale (from observations of standard
stars); the errors from the gaussian fits to the peaks in the profiles
were considerably smaller than this.

Note, finally, that although no H$_2$ emission was detected toward the
central continuum position (0\arcsec\ offset in each of the PV diagrams),
a spatially-compact feature is detected in [FeII].  In an extracted
spectrum (not shown) this feature peaks at $\sim -2$\kms, i.e.  roughly
the nominal systemic velocity based on the blue- and red-shifted
velocities seen in the lobes in H$_2$ and [FeII].  At only $\sim 24$\kms\
wide FWHM, this feature is also considerably narrower than the [FeII]
profiles in the lobes, being only moderately broader than the instrumental
profile.


\subsection{IRAS\,17423-1755}   
 
Optical images of the bipolar PPN \irasb\ (Hen\,3-1475) reveal a
spectacular S-shaped distribution of point-symmetric knots along the
axis of the nebula, which extend over $\sim$17\arcsec\ (Borkowski,
Blondin \& Hartigan 1997; Riera et al. 2003).  Like M\,1-92, this
chain of knots is enveloped in a diffuse, dumbell-shaped, bipolar
nebula.  From optical spectroscopy, velocities for the fast wind in
the lobes of the nebula of $\sim 1000$\kms\ have been inferred, and
shocks are once again thought to play a major role in the excitation
of the optical line-emission features (e.g. Riera et al. 2003).  Riera
et al. suggest that \irasb\ may have evolved from a relatively high
mass progenitor.  VLA observations do reveal a compact radio-continuum
source near the central star (Knapp et al. 1995), so the central
object may already be hot enough to produce an ionised region.

\begin{figure*} 

\epsfxsize=15.0cm
\epsfbox[15 210 600 580]{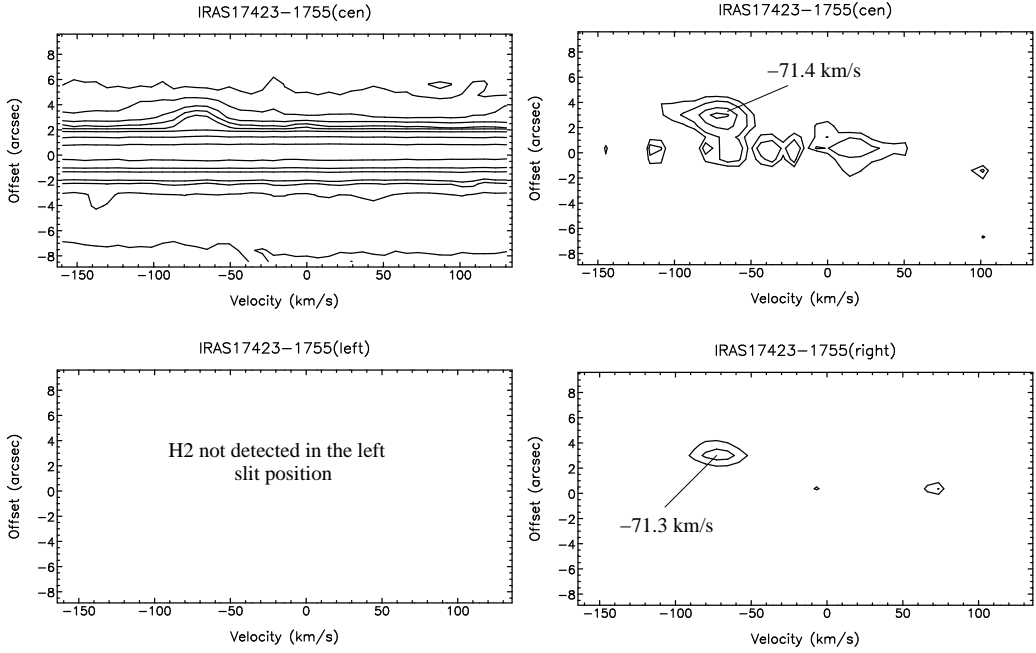}

\caption{PV diagrams showing the H$_2$ 2.122\mic\ emission in  \irasb\
(see Figure 2 for details). The top of each spectral
image is orientated roughly to the north-west (p.a. $\sim -45\degr$);
the ``left'' and ``right'' spectra were offset 1.0\arcsec\ to the
north-east and south-west, respectively. 
The contour levels measure 0.02,0.04,0.08,0.12,0.16,0.32 Jy
per arcsec along the slit in all four plots. }
\label{17423con}
\end{figure*}

H$_2$ observations at three positions across the width of \irasb\ are
shown in Figure~\ref{17423con}.  The outer slits coincide with the
edges of the diffuse lobes of scattered light, while the central slit
covers many (though not all) of the optical knots along the central
axis. Even so, H$_2$ line emission was only detected in the
north-western outflow lobe (positive offsets in
Figure~\ref{17423con}), probably toward the bright optical knot NW1a
(Riera et al. 2003).  Our continuum-subtracted spectral images do not
reveal emission coincident with the central star, although the shot
noise associated with the bright continuum is much stronger than in
the bipolar lobes.

%
%

The H$_2$ emission in the north-western lobe peaks at an offset of
$\sim$3\arcsec ; it is also strongest in our ``right'' slit position,
which corresponds to the south-western edge of this lobe.  The
emission feature peaks at a highly blue-shifted velocity ($V_{\rm LSR}
\sim -70$\kms ).  This is not too surprising when one considers 
that \irasb\ is probably the most obliquely-orientated source in our
target list (Table 1).  The line profile is narrow (FWHM $\sim 18$\kms
) though there is also possibly a weak wing extended to higher
blue-shifted velocities.


%
\begin{figure*} 

\epsfxsize=15.0cm
\epsfbox[15 210 600 580] {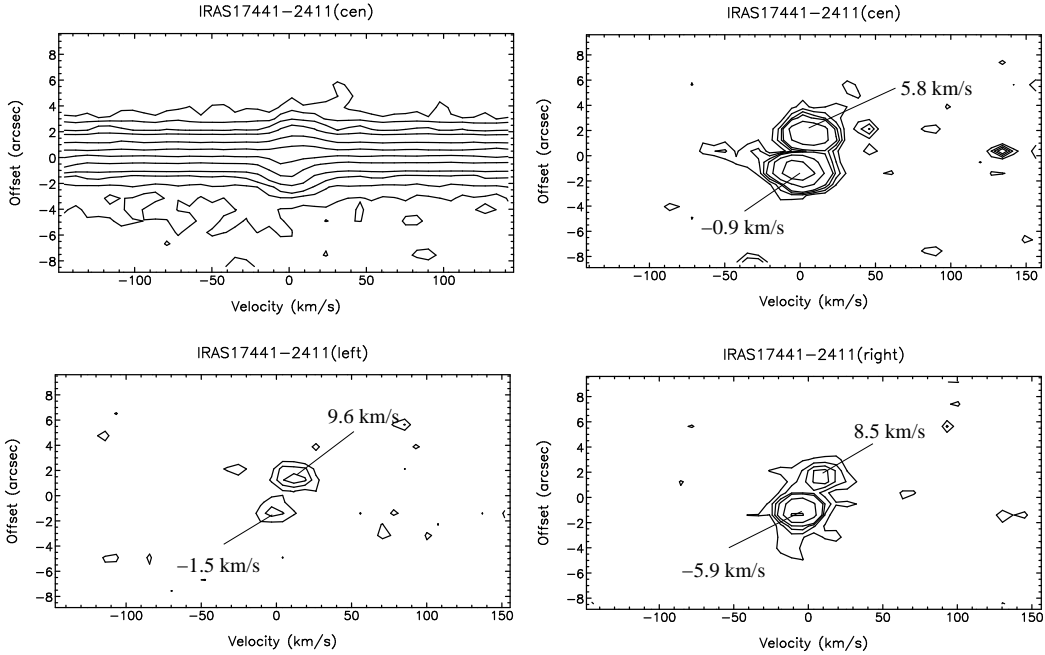}

\caption{PV diagrams showing the H$_2$ 2.122\mic\ emission in  \irasc\
(see Figure 2 for details).  The top of each spectral
image is orientated roughly to the north (p.a. $\sim 17\degr$);
the ``left'' and ``right''
spectra were offset 1.2\arcsec\ to the east and 0.8\arcsec\ to the west,
respectively. The contour levels measure
0.0025,0.005,0.0075,0.01,0.02,0.04,0.08 Jy per arcsec along
the slit in all four plots. }
\label{17441con}
\end{figure*} 
%

%
\begin{figure*} 

\epsfxsize=15.0cm
\epsfbox[15 210 600 580] {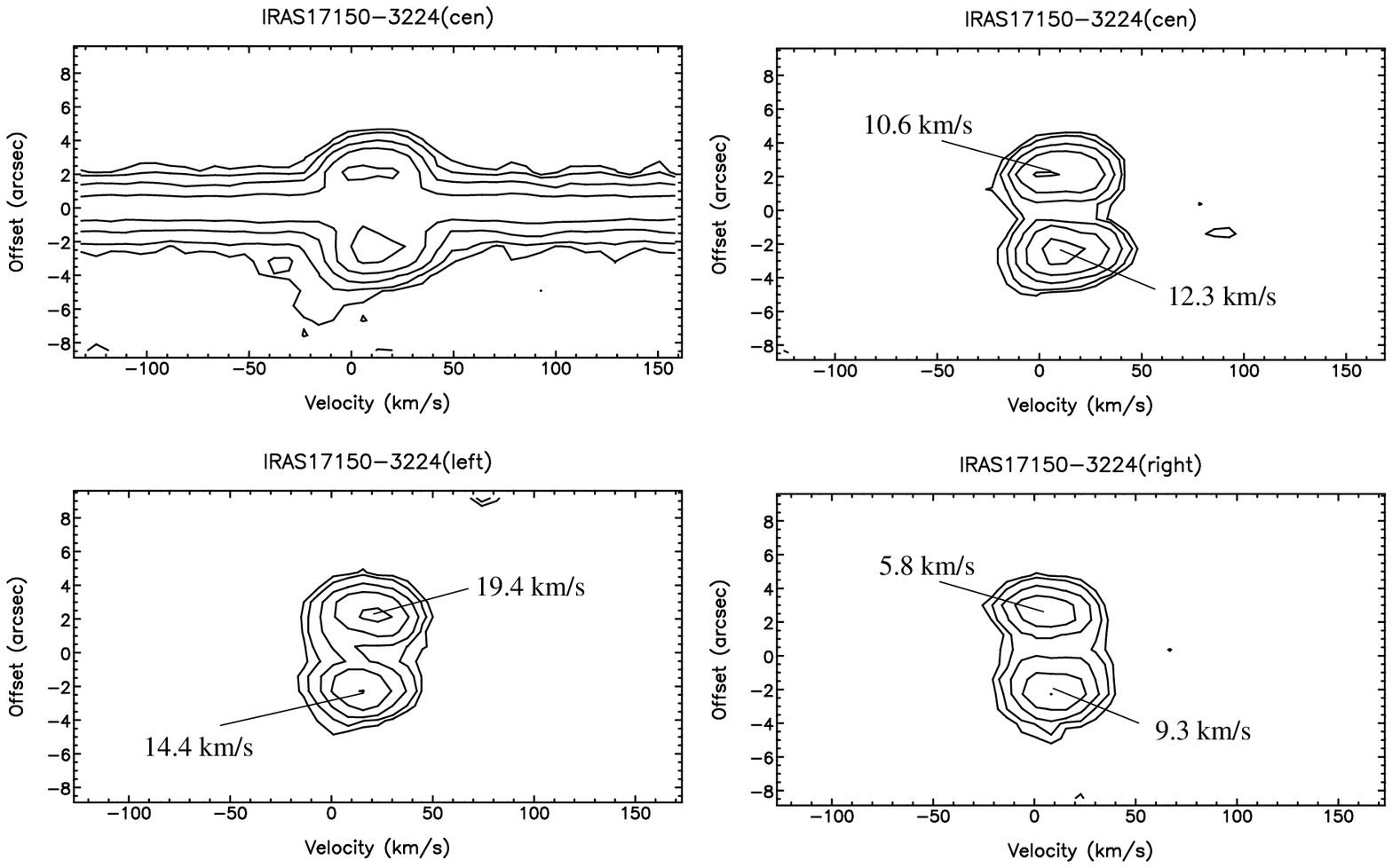}

\caption{PV diagrams showing the H$_2$ 2.122\mic\ emission in \irasa\
(see Figure 2 for details).  The top of each spectral image is
orientated roughly to the north-west (p.a. $\sim -55\degr$); the
``left'' and ``right'' spectra were offset 0.8\arcsec\ to the
north-east and south-west, respectively.  The contour levels measure
0.0015,0.0025,0.005,0.01,0.02,0.04,0.08,0.16,0.32 Jy per arcsec along
the slit in all four plots. }
\label{17150con}
\end{figure*} 
%

\subsection{IRAS\,17441-2411}   
 
With respect to angular size, aspect ratio and overall structure,
\irasc\ (AFGL~5385) is very similar to \irasa\ (discussed below).
Both targets are also associated with late-type central sources.  In
{\em HST} images \irasc\ comprises two bulbous lobes superimposed onto
a series of concentric arcs, which may be associated with
spherically-symmetric AGB mass loss (Su et el. 1998; 2003).  The lobes
themselves are not uniformly illuminated, instead exhibiting an ``S''
shaped geometry (Ueta et al. 2000). \irasc\ is probably orientated
close to the plane of the sky, since the object has a well-defined
dark lane across its waist which obscures the central star.  The star
is itself extremely red; V-K$>$11.5 (Su et al. 2003).

Weintraub et al. (1998) report the detection of weak H$_2$ emission
toward the core region in \irasc\ (Davis et al. 2003b failed to
detect H$_2$ in this source, probably because of the increased shot
noise in their lower-spectral-resolution observations).  Weintraub et
al. postulate that the H$_2$ emission may be associated with an
expanding equatorial torus.  However, their slit was orientated
east-west, rather than along the almost north-south bipolar lobes.

In Figure~\ref{17441con} we again show continuum-subtracted H$_2$
spectral images at three slit positions across the width of the
nebula.  As with the other PPN, the H$_2$ emission is confined to
discrete knots in the bipolar lobes; in \irasc\ these are offset
approximately 1\arcsec -2\arcsec\ from the central continuum peak in
our PV diagrams.  In optical images the lobes appear to be somewhat
limb-brightened.  In H$_2$ we detect the strongest line emission along
the central axis, although notably the northern, red-shifted knot
(positive offsets in Figure~\ref{17441con}) is brighter in the
``left'' slit than the emission feature in the counter-lobe, with the
inverse seen in the ``right''slit.  Note also that the velocity
difference between the blue and red lobes is marginally larger in the
left and right slits than it is along the central slit.  Curiously,
this implies higher radial velocities along the edges of the bipolar
lobes than along the central axis.  Higher spatial- and
spectral-resolution observations would be useful to confirm this
finding.

Overall, the low radial velocities in H$_2$ support the idea that the
system is orientated close to the plane of the sky.  Emission lines in
H$_2$ spectra extracted from the two bipolar lobes are again very
narrow; Gaussian fits yield FWHM values of $\sim 25$\kms .


\subsection{IRAS\,17150-3224 }   
 
\irasa\ (AFGL\,6815) also appears to be orientated close to the plane of
the sky, with its equatorial torus viewed nearly edge-on (Kwok, Su \&
Hrivnak 1998).  The central G-type star is largely obscured at optical
wavelengths and is highly reddened (V-K$>$11.0; Su et al. 2003).  Like
\irasc, in optical and near-IR {\em HST} images a series of concentric
arcs are superimposed onto the bipolar lobes; these may again be
associated with spherically-symmetric mass loss from the AGB phase (Kwok
et al. 1998; Su et al. 2003). Overall, the angular extent of the PPN,
measured in the optical across both bipolar lobes, is about 8\arcsec
-10\arcsec , although fainter rings of emission extend radially further
outward.

H$_2$ emission was detected by Weintraub et al. (1998) and Davis et al.
(2003b) in both bipolar lobes.  Our PV diagrams in Figure~\ref{17150con}
reveal the kinematics of these emission features.  The plots show that the
emission extends across the width of the nebula, and that once
again it is confined to knots in the two bipolar lobes. The knots are 
offset by $\sim$2\arcsec\ from the central continuum.  The emission is
therefore excited roughly half-way along the optical bipolar lobes, rather
than at the ends of the lobes.

Much lower radial velocities are recorded in \irasa\ ($V_{\rm LSR} \sim$
10 -- 12~\kms , FWHM $\sim 35$\kms ) than were seen in some of the other
targets, as one would expect for a system orientated close to the plane of
the sky.  (The same peak velocity was measured by Weintraub et al. (1998)
in their $R \sim 65,000$ \htwo\ spectrum obtained with an east-west slit.)

\begin{table*}
\caption{\htwo\ and \brg\ line-profile fitting results from the spectra 
in Figure~\ref{h2brg-spec}.  Peak velocities and FWHM widths
were obtained from either Lorentzian or Gaussian fits.  The former
gave a better fit to the \brg\ data; the latter more closely matched
the H$_2$ profiles.  }
        \begin{tabular}{lcccccc}
        \hline  
Target      &   H$_2$ (blue)  &  H$_2$ (blue)   &   H$_2$ (red)   &  H$_2$ (red)   &   Br$\gamma$    &   Br$\gamma$     \\
            & $V_{\rm peak}$  &     FWHM        & $V_{\rm peak}$  &     FWHM       &  $V_{\rm peak}$ &     FWHM         \\
            &  (km s$^{-1}$)  & (km s$^{-1}$)   &  (km s$^{-1}$)  & (km s$^{-1}$)  &  (km s$^{-1}$)  & (km s$^{-1}$)    \\

        \noalign{\smallskip}
        \hline
        \noalign{\smallskip}

\irasd      &   14.5   &  27.6     &  14.7  &  29.7    &    26.8       &  25.8   \\
M~1-92      &  -33.8   &  43.3     &  28.7  &  37.2    &   -11.4       &  87.4   \\
\irasb      &  -69.7   &  17.7     &   --   &   --     &    43.5       &  83.7   \\
\irasc      &   -1.0   &  25.9     &  5.9   &  26.4    &     4.9$^*$   &  65.3$^*$   \\
\irasa      &  10.9    &  36.3     &  12.3  &  34.6    &    20.4$^*$   &  93.1$^*$  \\ 

        \noalign{\smallskip}
        \hline
        \noalign{\smallskip}
        \end{tabular}

 \begin{list}{}{}
 \item[$^*$]HI lines seen in absorption 
 \end{list}
  \label{tab2}
\end{table*}


\subsection{H$_2$ non-detections in IRAS\,09371+1212 and IRAS\,17106-3046} 
 
Two other PPN were also observed at 2.12\mic , IRAS\,09371+1212
(Frosty Leo) and IRAS\,17106-3046 (see Table 1).  Both targets appear
to be bipolar PPN, much like those described above, and both are
associated with collimated, knotty outflows (e.g. Langill, Kwok \&
Hrivnak 1994; Sahai et al. 2000; Kwok, Hrivnak \& Su 2000). Even so,
\htwo\ line emission was not detected along their central bipolar
axes.

The angular extent of IRAS\,09371+1212 on the sky is considerably
larger than the other targets in our sample, so it is certainly
possible that our chosen slit position angle missed line-emission
features along the walls of the nebula.  Our slit was aligned through
bright, optical ansae at the ends of the bipolar lobes (labeled A$_N$
and A$_S$ by Sahai et al. (2000)), features which tend to be H$_2$
line emitters in other PPN.  Also, the outflow itself may have changed
direction in recent years, as evidenced by the multiple, miss-aligned
optical features observed in the lobes of this nebula (Sahai et
al. 2000).

IRAS\,17106-3046 has a much smaller angular extent on the sky, though
it is still comparable in size to \irasb\ and \irasd, PPN where H$_2$
was detected and spatially resolved.

Inclination angle does not seem to be a factor in
whether H$_2$ is detected or not.  In the other targets discussed
above, H$_2$ is detected in sources with a range of inclination angles
(see Table 1).

Both targets certainly show evidence for collimated flows, and the
knots and ansae in IRAS\,09371+1212 in particular indicate the
presence of shocks.  Shock velocities should be high enough to excite
H$_2$ into emission; the 1-0S(1) line targeted here can be excited in
shocks with velocities of $\le 10$\kms\ (e.g. Smith 1991).  The lack
of detectable emission must therefore be related to a lack of dense,
molecular gas in the polar lobes.

Our H$_2$ non-detections in IRAS\,09371+1212 and IRAS\,17106-3046 may be
related to their progenitor mass and/or evolutionary state. Weintraub et
al (1998), Garc\'{\i}a-Hern\'andez et al. (2002) and Davis et al. (2003b)
all find that H$_2$ is not detected in targets with spectral types later
than about G2.  Our non-detection in IRAS\,09371+1212, which has a K7II
central object (Mauron, Le Borgne \& Picquette 1989), would support this
hypothesis.  The spectral type of IRAS\,17106-3046 is not known.

%
\begin{figure} 

 \epsfxsize=8.0cm
 \epsfbox{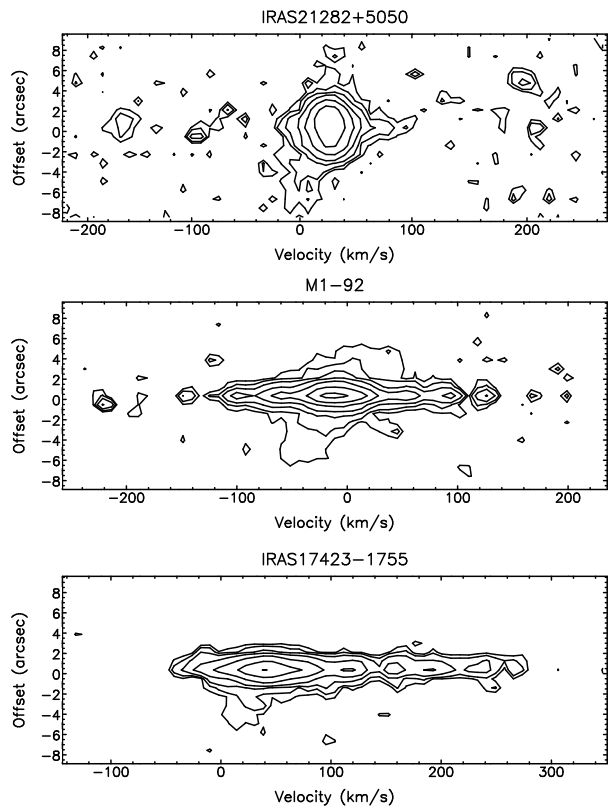}

\caption{PV diagrams showing the \brg\ emission in three of the five
targets observed (\brg\ was seen only in absorption in \irasa\ and
\irasc ). The continuum emission (at zero offset in the y-axis) has been fitted
and removed.  In all three plots the contour levels start at
0.025 Jy arcsec$^{-1}$ and increase in multiples of 2.}

\label{brg-cont}
\end{figure} 
%

\subsection{\brg\ observations} 

\brg\ observations were acquired for the central slit position, using the
same slit position angle, in each of the five targets where H$_2$ was
detected.  In two of the five sources -- the two late, G-type sources in
our sample -- \brg\ is seen in absorption.  This is typically the case in
F or G-type PPN (e.g. Hrivnak, Kwok \& Geballe 1994).  In the ``more
evolved'' B-type PPN and the ``young PN'' \irasd , \brg\ is strong and
seen in emission, indicating the onset of ionisation.  M1-92, \irasb\ and
\irasd\ are probably on the brink of the transition from PPN to PN (Kwok
1993). In these three targets the emission is largely confined to the
source continuum position (Figure~\ref{brg-cont}), although weak,
blue-shifted emission, presumably associated with the inner regions of the
post-AGB wind, is also seen in each case.

%
\begin{figure} 

 \epsfxsize=6.0cm
 \epsfbox{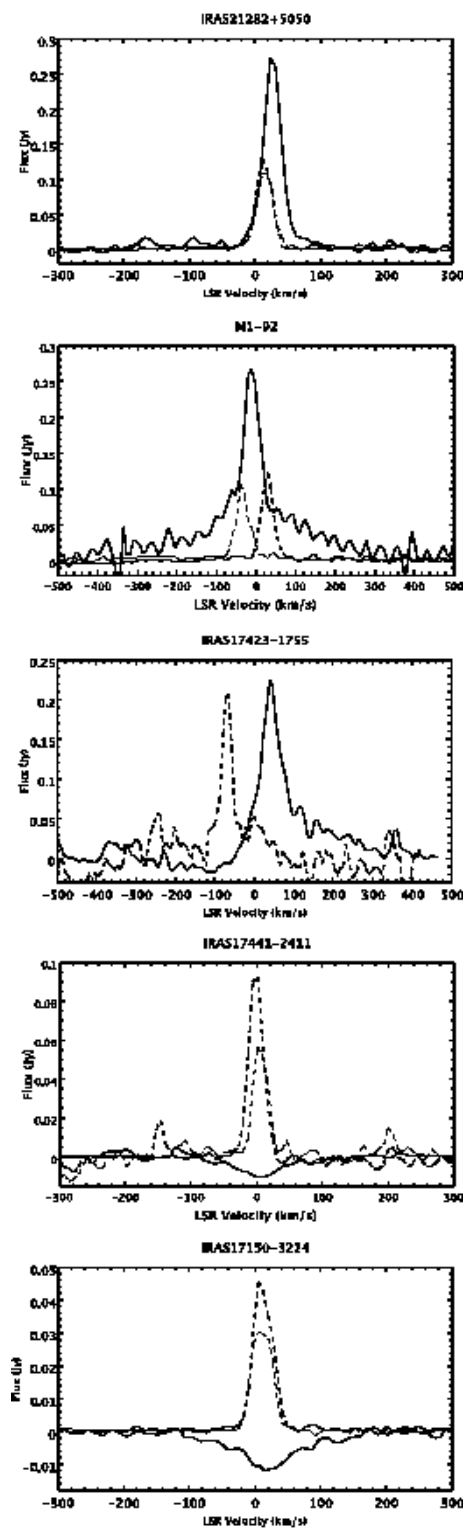}

\caption{\htwo\ and \brg\ spectra: H$_2$ spectra have been extracted
over a 2.7\arcsec -wide strip $\sim 2$\arcsec\ above and below the
source continuum and are drawn with dashed lines (in \irasb\ H$_2$
emission was only detected in one bipolar lobe); the Br$\gamma$
spectra are extracted over the same area though these are centred on the
source continuum position. The H$_2$
spectra for M1-92 have again been scaled by a factor of 0.1.}
\label{h2brg-spec}
\end{figure} 
%

In Figure~\ref{h2brg-spec} we compare \brg\ profiles to \htwo\
profiles, the latter being extracted from the bipolar lobes of each
PPN.  The central slit images were used for the H$_2$ spectra.  The
results of Gaussian (H$_2$) or Lorentzian (\brg ) fits to these lines
are listed in Table 2. The \brg\ profiles generally peak at low radial
velocities, velocities that are comparable (to within $\pm$10~\kms )
to systemic velocities derived from CO observations of the molecular
gas associated with each system (Bujarrabal et al. 20001 and
references therein).  Considering the targets individually;

\begin{itemize}

\item{The two H$_2$ profiles from the young O-type PN \irasd\ peak at
essentially the same low velocity, as one would expect for a system
orientated close to the plane of the sky.  The \brg\ profile peaks at
a similarly low velocity, and is notably much narrower than the \brg\
profiles observed toward the other sources (Table 2). }

\item{In the B-type PPN M1-92, the \brg\ emission peaks midway between
the two H$_2$ emission lines.  Here, the HI recombination line is
probably centred at the systemic velocity, while the H$_2$ lines
reflect the blue- and red-shifted gas velocities in the bipolar
lobes.}

\item{In the B-type PPN \irasb\ there is a considerable shift between
the H$_2$ and \brg\ peak velocities.  Again, the \brg\ line peaks at
the systemic velocity (as reported by Riera et al. 2003), while the
H$_2$ traces the high-velocity blue-shifted outflow lobe.  The
red-shifted lobe may be obscured by a circumstellar torus inclined
toward the plane of the sky.  The large velocity shift ($\sim
130$~\kms ) between the H$_2$ and \brg\ profiles would likewise be
caused by the large inclination angle of the system.}

\item{In the G-type PPN \irasc\ and \irasa\ the \brg\ absorption features
again probably mark the systemic LSR velocity in each system. The H$_2$
profiles in the bipolar lobes in both targets peak at roughly the same
radial velocity, as one would expect for nebulae orientated close to the
plane of the sky.}

\end{itemize}

In four of the five sources (the PPN) the extracted \brg\ profiles are
considerably broader than the H$_2$ profiles, by a factor of $\sim 2$
(Table 2).  The \brg\ emission must trace ionised regions within the fast
post-AGB wind, while the H$_2$ instead reflects the shock velocity at the
interface between the ionised wind and the slow-moving AGB envelope.  
Note in particular the broad blue and red-shifted wings superimposed onto
the otherwise Lorentzian profile in M~1-92.  These wings reach velocities
of $\pm 200$\kms\ (at the 10\% peak intensity level).  For an inclination
angle of $\sim 35\degr $ this indicates a post-AGB ionised wind speed of
at least 350\kms . 

In contrast with the other sources, the \brg\ profile toward the young
PN \irasd\ is very narrow, being comparable in width to the H$_2$
profiles.  We also see no evidence for associated line wings.
Together, the H$_2$ and \brg\ data support Meixner et al (1997)'s
claim that the bipolar lobes are orientated close to the plane.  Our
data also suggest that the HI recombination line emission must be
excited in a poloidal rather than equatorial wind, and that this wind
must be highly collimated, perhaps by the torus reported by Meixner et
al., since any lateral expansion would serve to broaden the line
profile.

%
\section{Further Discussion}
%

\subsection{\brg\ and H$_2$ as tracers of the fast wind}

\brg\ and H$_2$ clearly trace different components in the outflows
from the sources considered in this paper.  \brg\ emission is produced
close to the central continuum source, the line profiles are generally
broad though they peak close to the systemic velocity, and the \brg\
flux in each system rapidly decreases with distance along the bipolar
axes.  H$_2$, on the other hand, is predominantly excited in the
bipolar lobes in each PPN and appears to be confined to shocks within
each flow (we detect molecular line emission toward the
continuum-peak in only one system, \irasd ). H$_2$ profiles are
narrow, though their peak velocities are blue- and red-shifted in
bipolar nebulae that are inclined out of the plane of the sky (note in
particular \irasb ).  We will discuss the origin of the H$_2$ emission
in subsequent sections.  However, we first consider possible
excitation mechanisms for the observed \brg\ emission.

Detectable \brg\ recombination-line emission requires high gas
densities or an intense ionising flux.  The former is produced in
post-shock gas; the latter exists close to intermediate or high-mass
stars.  High-resolution imaging and kinematic studies of PPN show them
to be associated with outflows and shock phenomena. PPN, as precursors
to PN, are also potentially associated with compact, photo-ionised regions.
Either environment may therefore be a source of the observed \brg\
emission.

Atomic hydrogen is collisionally-ionised in shocks, provided the
post-shock gas temperature exceeds $\sim 4\times 10^4$K, i.e. for
shock velocities greater than about 50~\kms .  Radiative recombination
then leads to bright Lyman and Balmer emission.  The higher-energy
Bracket lines will be much weaker, of the order of 0.01--0.03 times
the H$\beta$ intensity for shock velocities up to 150~\kms\ and gas
densities in the range $10^3-10^7$\,cm$^{-3}$ (Hollenbach \& McKee
1989).  Indeed, for a shock velocity of $\sim$100-150\kms\ and a
pre-shock gas density of $\sim 10^4$~cm$^{-3}$, Hollenbach \& McKee
predict a \brg\ line intensity of the order of
$10^{-3.7}$~erg~cm$^{-2}$~s$^{-1}$~sr$^{-1}$.  For a shock covering an
area of 1\arcsec\ ($2.35\times10^{-11}$~sr) on the sky, and for a line
width of $\sim 80$\,\kms\ (Table 2), this is equivalent to a \brg\
line flux density of about 0.01~Jy.  This is an order of magnitude
lower than the observed flux levels in the two early-type PPN and the
young PN, \irasd\ (Figure~\ref{h2brg-spec}).  Much higher gas
densities, of the order of 10$^5$--10$^6$~cm$^{-3}$, would be needed
to produce the observed line strengths.

In an ionised region at the base of a fast wind,
Br$\gamma$ will again be relatively weak in comparison to Lyman and
Balmer lines, $\sim$0.03 times the H$\beta$ intensity for an
electron temperature and density of $\sim 10^{4}$~K and $\sim
10^{4}$~cm$^{-3}$ (under Case B [optically thick to Ly$\alpha$]
recombination theory; Hummer \& Storey 1987).  The total integrated
intensity in \brg\ from an optically thin PPN nebula of radius $R$ at
a distance $D$ will be (Kwok 2000): \\

\noindent
\begin{math}
  I_{Br\gamma } = 3.41\times10^{-27} n_e n_p 
      \left( \frac{R^3 \epsilon}{3 D^2} \right) 
           \hspace*{0.2cm} {\rm erg \ cm}^{-2}{\rm s}^{-1}.    
\end{math}  \\

\noindent For $D=2$\,kpc, $R=0.01$\,pc ($\sim$2\arcsec ), $n_{\rm e}
\sim n_{\rm p} = 10^4$\,cm$^{-3}$ and a filling factor, $\epsilon$, of
unity, the total
\brg\ flux will be $\sim0.9\times 10^{-13}$~erg~cm$^{-2}$~s$^{-1}$. For 
a line width of $\sim 80$\kms (Table 2), the flux density will be
$\sim 0.2$~Jy, comparable to the values in our extracted
spectra (Figure~\ref{h2brg-spec}).

Thus, from the observed line intensities, a photo-ionised
nebula close to the central object seems to be the most likely origin
for the \brg\ emission.  The spatial distribution of \brg\ along the
slit also tends to support recombination at the base of the outflow,
rather than in shocks between the fast wind and the AGB envelope,
since these shocks, traced in H$_2$ in each system, are at a
resolvable distance from the central source and yet the \brg\ emission
is largely confined to regions near the central source.  The low peak
velocities though broad line widths in the \brg\ spectra also support
recombination in a fast, possible poorly collimated or even spherical
wind close to the source.

Finally, we mention that the observations of Guerrero et al. (2000)
and Davis et al (2003b) and the models of Natta \& Hollenbach (1998)
indicate that the \htwo /\brg\ ratio is highest in PN with large,
well-defined rings and bright central stars, i.e. in evolved and/or
intermediate-to-high mass sources.  The ratio seems to be lower in
``less-evolved'' PN.  When we consider PPN, particularly those with
late-type central stars like \irasc\ and \irasa , where thermal and
probably shock-excitation dominates the production of H$_2$ line
emission, the \htwo /\brg\ ratio is again high.


\subsection{The ``three wind'' model}

Three phases of mass loss have been invoked to explain the multiple
shells observed in some PN (e.g. Frank 1994).  In this scenario the
phase of slow, spherical AGB mass loss ({\em
\.M}$\sim10^{-7}$--$10^{-6}$~\Msolar\ yr$^{-1}$; $V\sim10$~\kms ) is
followed by a slow though more massive ``superwind'' ({\em
\.M}$\sim10^{-5}$--$10^{-3}$~\Msolar\ yr$^{-1}$; $V\sim10$~\kms ).  A
diffuse, fast wind ({\em \.M}$\sim10^{-8}$~\Msolar\ yr$^{-1}$;
$V\sim1000$~\kms ) follows a few thousand years after the superwind
has been terminated.  The model is probably a simplification of a
continually changing post-AGB mass-loss rate, and in Frank (1994)'s
treatment only the ``preionisation stage'' is relevant to
PPN. Nevertheless, we investigate briefly whether our near-IR spectra
can be used to distinguish between the traditional ``Two-Wind'' model
of Kwok (1993) and a more complex scenario.  Is an additional
superwind required to explain the H$_2$ observations?

%
\begin{figure*} 

   \plotfiddle{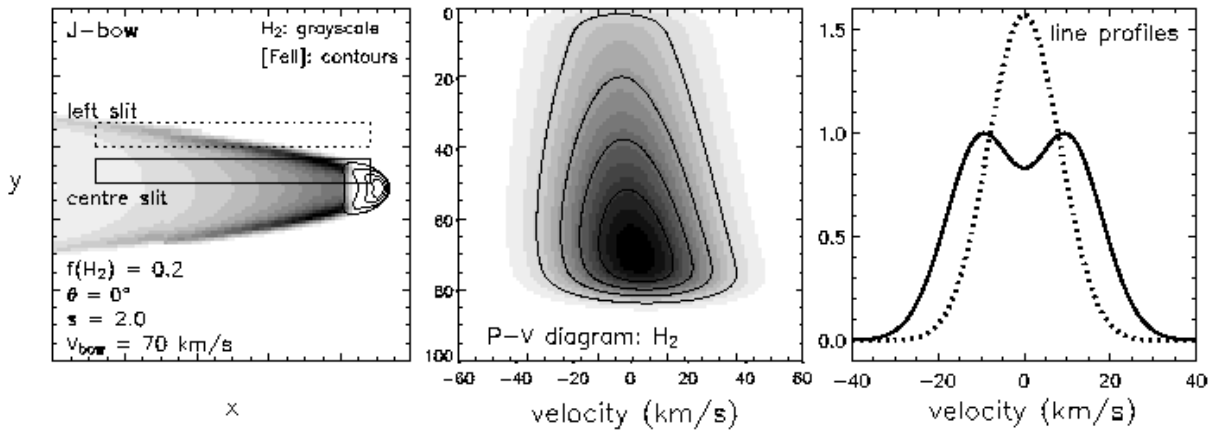}{60mm}{-90}{65}{65}{-270}{290}

\caption{A bow shock model for the H$_2$ emission from \irasd . The
H$_2$ profiles in the third panel can be compared to the extracted
spectra in Figure~\ref{21282spec}.  In the model, a parabolic bow of
speed 70\,km\,s$^{-1}$ advances into a uniform medium of hydrogen
density $n_{\rm H} = 10^4$\,cm$^{-3}$ (of which 60\% is in atomic
form) and Alfv\'en speed 1~km~s$^{-1}$ ($B \sim 50$~$\mu$G), with the
magnetic field parallel to the bow motion. The size (semi-latus
rectum) of the bow is 2\,$\times$\,10$^{16}$\,cm ($\sim$0.5\arcsec\ at
3~kpc). The line profiles in the third panel correspond to the
space-integrated emission from the two slits shown in the first
panel. The PV diagram in the middle panel has been calculated by
integration through the entire transverse direction.  The line
profiles were convolved with a Gaussian of velocity dispersion
7.2~km~s$^{-1}$ and the PV diagram with 15~km~s$^{-1}$.  In the left
panel the star would be to the left; in the middle panel the star
would be to the top.}

\label{IRAS21282model}
\end{figure*} 
%

The density discontinuity between a slow superwind and an equally-slow
AGB envelope drives a boundary shock (Frank 1994) which could produce
strong H$_2$ line emission.  However, it would not generate the high
radial velocities seen in our H$_2$ spectra (velocities approaching
50-100\kms\ in some case; Table 2) since neither the superwind nor the
AGB wind expand at high speed.  So the observed H$_2$
emission must not derive from a Superwind/AGB wind boundary. Instead,
a fast wind impacting the back of either a superwind or an AGB wind
is needed to produce the observed high velocities.

The boundary layer between a diffuse, fast wind and a dense, neutral
superwind or AGB wind would comprise a weak, outer/forward shock and a
strong, inner/reverse shock.  The latter, with $V_{\rm shock} \sim
1000$\kms , decelerates and heats the fast wind material to very high
temperatures ($T \sim 10^7$K).  The velocity of the forward shock would be
much lower, comparable to though somewhat higher than the velocity of the
AGB wind, depending on the ratio of the fast wind density and the AGB or
superwind density. For example, a fast wind velocity ($V_{\rm fast}$) of
1000\kms\ and a density ratio between the fast wind and AGB wind
($\rho_{\rm AGB}$/$\rho_{\rm fast}$) of 100 would result in an outer shock
velocity, given by \\
  
\noindent
\begin{math}
V_{\rm shock} \sim 
         V_{\rm fast}/ [1 + (\rho_{\rm AGB}/\rho_{\rm fast})^{1/2}],
\end{math}\\

\noindent of $\sim90$~\kms . By comparison, a density ratio of 10,000 between
the same fast wind and a much denser superwind would yield a much
lower shock velocity, of the order of 10\,\kms.  In both cases,
because the shocked H$_2$ would more-or-less move with the shock
front, the peak H$_2$ velocity in observed spectra (and the proper
motions of H$_2$ emission knots), when corrected for the flow
inclination angle, should be comparable to these shock velocities.
Since moderately high H$_2$ velocities have been observed, our
observations tend to support a lower density ratio and so, at least
for the H$_2$ emission regions considered here, a third, dense
superwind component may not be required.


\subsection{H$_2$ excitation in bow shocks}

The morphologies and excitation of molecular hydrogen features in PPN,
when combined with the kinematic data presented here and elsewhere
(e.g. Weinbtraub et al. 1998), strongly support excitation in curved
or conical shocks.  Models of similar ``bow'' shocks have been invoked
to explain line profiles and excitation maps in HH jets and molecular
outflows from young stars, most recently the molecular bow shocks
observed in embedded systems (e.g. Smith 1991, 1994; Davis et
al. 1999; Smith, Khanzadyan \& Davis 2003; O'Connell et al. 2004).  In
these papers the effects of varying orientation angle, bow shock
speed, bow shape, pre-shock density, and the Alfv\'en speed and
magnetic field orientation are investigated and discussed in detail.
Here we present only the models which best fit the kinematic data
obtained for two systems: \irasd , where complex and double-peaked
profiles are observed, and M\,1-92, where H$_2$ and [FeII] emission is
detected. In both cases we find that a J-type bow shock fits the data
reasonably well; in other words, C-shocks and therefore strongly
magnetised winds are not required.

Wherever possible we use parameters derived from other
observations to constrain the models.

\subsubsection{\irasd } 

\noindent The bipolar lobes of the \irasd\ nebula appear to be orientated
close to the plane of the sky.  In H$_2$ we observe emission along
the flow axis, but also toward the central continuum peak, where
double-peaked profiles with a peak-to-peak separation of 22~\kms\ are
observed (Figure~\ref{21282spec}).  In Section~3.1. we discuss whether
the emission toward the continuum could be excited in an equatorial
wind. We have attempted to model such an equatorial outflow with bow
shocks propagating along the line of sight. This might be considered
consistent with the details displayed in Figure~\ref{21282spec} and
discussed in Section~3.1; specifically, the double-peaked 1--0\,S(1)
profile plus single peaked profiles from the left and right-hand
spectra. However, after detailed modeling we were unable to reproduce
the peak shifts. This is because bow shocks tend to deflect material
away from the shock surface rather than plough it up. Even with a slow
non-dissociative bow shock, most of the emission arises from the wings
where the emission peak is shifted by only a few km~s$^{-1}$.

%
\begin{figure*} 

   \plotfiddle{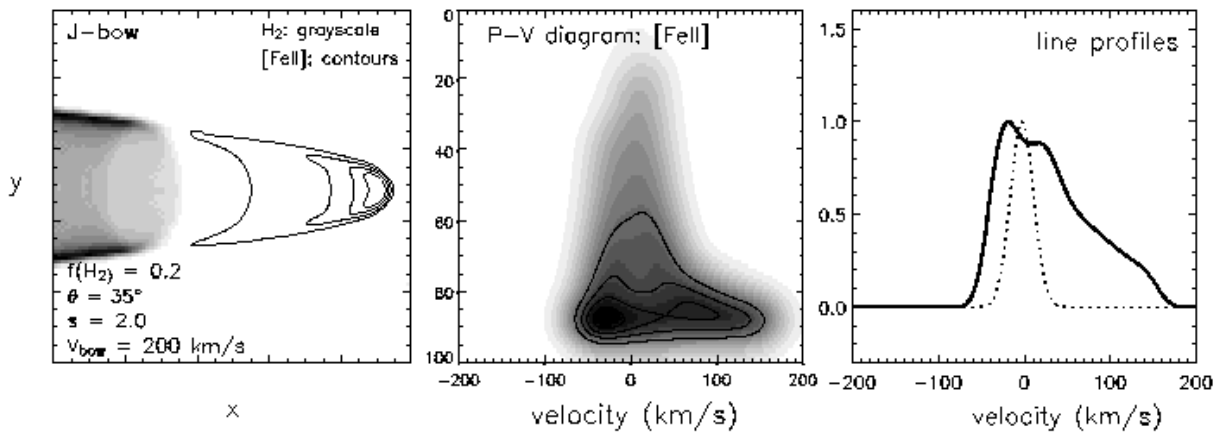}{60mm}{-90}{65}{65}{-270}{290}

\caption{A bow shock model for the H$_2$ and [FeII] emission from
M~1-92. The [FeII] PV diagram and H$_2$ and [FeII] line
profiles can be compared to the observed red-shifted components shown
in Figures \ref{m192con2} and \ref{m192h2fe-spec}.  In the model, a
parabolic bow of speed 200~\kms\ advances into a uniform medium with
hydrogen density $n_{\rm H} = 10^4$\,cm$^{-3}$ (60\% in atomic form)
and Alfv\'en speed 1~km~s$^{-1}$ ($B \sim 50 \mu$G), where the field
is parallel to the bow motion. The size (semi-latus rectum) of the bow
is 2\,$\times$\,10$^{16}$\,cm ($\sim$0.6\arcsec\ at 2.5~kpc). In the
centre panel, the [FeII] PV diagram has been calculated by integration
through the entire transverse direction.  In the right-hand panel, the
[FeII] (full line) and H$_2$ (dotted line) profiles correspond to the
space-integrated emission from horizontal slits passing through the
bow apex and the bow wing, respectively. The PV diagram and line
profiles have been convolved with a Gaussian of velocity dispersion
15~\kms\ and 12~\kms , respectively.  In the left panel the
star would be to the left; in the middle panel the star would be to
the top.}

\label{M192model}
\end{figure*} 
%

Instead, we more easily model the \irasd\ H$_2$ emission if it arises
from bow shocks moving in the plane of the sky, as displayed in the
left panel in Figure~\ref{IRAS21282model}. In this case, the
deflection from the bow flanks is found to be sufficient to generate a
twin-peaked H$_2$ profile from the central slit, with a peak-to-peak
separation consistent with the observations in Figure~\ref{21282spec};
a single-peaked profile is also produced in the bow wings,
corresponding to the left or right slit in Figure~\ref{21282spec}.
The overall line widths are also roughly consistent with the
observations, and the model profiles peak at the systemic velocity, as
is more-or-less observed to be the case.  The profiles have
been convolved with a Gaussian profile to represent turbulent motions
which will, in any case, be generated wherever there are curved shock
fronts.

In Figure~\ref{IRAS21282model} we also reproduce synthetic H$_2$ (and
[FeII]) images, for comparison with future high-resolution imaging,
and an integrated PV diagram, for comparison with the plots in
Figure~\ref{21282con}.  (Note that the model only yields emission from
one bow shock and therefore only one outflow lobe; in the observed PV
diagrams the emission is presumably excited in both lobes.)

The model employed solves the 1-D MHD equations at each
point on the bow surface. The cooling behind the shock and the motion
away from the bow surface is taken into account with the assumption
that the cooling length is short (but not negligible) in comparison to
the bow size.  The orientation derived from this model is
consistent with the direction of the bipolar outflow on larger scales
(Table 1). However, we cannot exclude other models given the present
limited dataset and modest resolution.

\subsubsection{M1-92} 

\noindent M~1-92 was the only PPN where we detected H$_2$ and
[FeII] emission: we therefore attempted to model both emission lines,
using bow shocks orientated at various angles to the line of sight.
We also varied the density, bow shape, field strength, molecular
fraction and bow speed. We obtained good agreement with the data for a
high-velocity bow shock propagating at 35$\pm$10$^\circ$ to the sky
plane, for a density exceeding 10$^3$\,cm$^{-3}$ and a bow shape $s
\sim 2.0 \pm 0.2$, where the bow shape is defined by $R \propto Z^s$ 
in cylindrical coordinates (e.g. Smith et al. 2003). The density,
velocity and inclination angle used are consistent with those derived
through modeling of atomic lines in the optical (Bujarrabal et
al. 1998a).

The model results are shown in Figure\,\ref{M192model}, where we again
display synthetic H$_2$ and [FeII] images of the bow structure.  As is
typically observed in HH bow shocks, the [FeII] emission is confined
to the apex of the bow, with the H$_2$ brightest in the
non-dissociating bow flanks.  Because the M~1-92 bow model is
advancing at a much higher speed than the \irasd\ bow, the [FeII]
emission is more extended and the H$_2$ is pushed further back into
the bow flanks (compare Figure\,\ref{M192model} with
Figure\,\ref{IRAS21282model}).  In high-resolution H$_2$ images of
M~1-92, the H$_2$ emission does seem to be confined to off-axis peaks
that could be the wings of a bow shock (Bujarrabal et al. 1998a; Davis
et al. 2003b).  From profiles extracted along the slit axis
(Section~3.2), we also find that the [FeII] peaks in the two lobes of
the bipolar nebula are slightly further apart than their H$_2$
counterparts, as would be expect from two oppositely-directed bow
shocks.

The predicted PV diagram for the [Fe\,II] data is shown in the second
panel of Figure~\ref{M192model}.  The plot displays a boot-shaped
structure comprising an extended red wing and faint, low-velocity
emission trailing back toward the source.  The overall shape of the
model PV diagram agrees quite well with the observed profile in
Figure~\ref{m192con2} for the southern emission lobe, although the
model does predict double-peaked profiles which are not observed.

Extracted H$_2$ (dotted line) and [FeII] (full line) spectra are shown
in the third panel in Figure~\ref{M192model}.  The model reproduces
the general structure of both the H$_2$ and [Fe\,II] lines; i.e. the
narrow, Gaussian H$_2$ profile from the bow wings together with the
broader, asymmetric [FeII] profile from the bow apex
(compare the dotted H$_2$ and full-line [FeII] model spectra in
Figure~\ref{M192model} with the observed spectra in
Figure~\ref{m192h2fe-spec}).

The extreme [Fe\,II] line wings, which extend out to $>300$~\kms , are
not reproduced by the model.  Moreover, the predicted H$_2$ and [FeII]
lines peak at or very near to the velocity of the ambient medium,
which was taken to be stationary.  The observed profiles are
red-shifted (and blue-shifted) by $35-45$~\kms\
(Figure~\ref{m192h2fe-spec}).  The observations imply that the
pre-shock medium is expanding with a radial speed of
$\sim$70~km~s$^{-1}$.  Such a fast-moving pre-shock medium ahead of
each bow would be sufficient to shift the model line profiles and PV
diagram by about 35~km~s$^{-1}$.

As with \irasd , the bow models do an adequate job of explaining the
observations, although additional high-spatial-resolution near-IR
imaging and spectroscsopy would again be useful to more tightly
constrain the theory.


\subsection{Rotation in a magneto-centrifugal disk wind}

Finally, since we have observed H$_2$ spectra at three positions
across the width of each PPN, we are potentially in a position to
search for signs of rotation.  Hints of jet rotation have recently
been identified on large scales in an outflow from a young star by
Davis et al. (2000), and investigated in more detail in HH jets by
Bacciotti et al. (2002) and Coffey et al. (2004).  In
magneto-centrifugal disk winds, the outflow is expected to maintain a
record of disk rotation, particularly close to the central star, so
their findings are not unexpected.

\irasa\ is the only source that shows possible signs of rotation.
This source is orientated close to the plane of the sky, is one of the
brightest H$_2$ emitters in our sample, and has emission features well
resolved from the central continuum.  In Figure~\ref{17150con} we see
a trend of increasing velocity from right to left across the width of
the outflow.  In other words, the south-western edges of both lobes
appear to be moving toward the observer, with the north-eastern edges
moving away from the observer, with a toroidal velocity of 3-7~\kms ,
depending on which lobe is considered. The trend is consistent across
all three slits, and is in the same sense in both lobes.  If we adopt
the slit offset (0.8\arcsec , or $3\times 10^{14}$~m at a distance of
2.4~kpc) as the radial distance, then an angular velocity of $\sim
2\times 10^{-11}$~rad s$^{-1}$ and an orbital period of 20,000\,years
is implied.

Disk wind models are reviewed by K\"onigl \& Pudritz (2002) and
discussed in the context of jet rotation by Bacciotti et al. (2000)
and in terms of PPN by Frank \& Blackman (2004).  Briefly, in a disk
wind material is ejected from the surface of the disk at a
``footprint'' radius, $r_{\rm o}$, and carried poloidally outward
along field lines, provided the base of the field lines are inclined
at less that 60\degr\ to the disk plane.  The point along the flow
where the poloidal flow velocity reaches the poloidal Alfv\'en speed
is known as the Alfv\'en surface; the flow radius at this point is
called the Alfv\'en radius.  The outflowing gas is accelerated only
between the disk surface and the Alfv\'en surface, where gas motions
can be approximated to solid body rotation.  In this region the
toroidal velocity is proportional to the radius; $v_{\phi}
\propto r$.  Beyond the Alfv\'en surface angular momentum is conserved
as the gas moves along the field lines; in this region $v_{\phi}
\propto r^{-1}$.  Consequently, at some point further downwind along 
the flow axis, the observed linear rotational velocity and radius,
$v_{\rm obs}$ and $r_{\rm obs}$, can be related to the rotational
velocity and radius at the Alfv\'en surface, $v_{\rm A}$ and $r_{\rm
A}$, and the Keplerian velocity and radius at the flow footprint,
$v_{\rm o}$ and $r_{\rm o}$, by:

\begin{equation}
  v_{\rm obs} r_{\rm obs} \sim  v_{\rm A} r_{\rm A} 
      \sim v_{\rm o} (r_{\rm A}/r_{\rm o}) r_{\rm A}.
\end{equation}

\noindent We assume that the flow does not interact with (and be
appreciably decelerated by) the ambient medium: hence, measurements as
close as possible to the disk are desirable.  For a ``cold'' flow the
maximum poloidal velocity along the magnetic field lines will be
equivalent to the outflow velocity.  This is given by:

\begin{equation}
 v_{\rm flow} \sim 2^{1/2} v_{\rm o}(r_{\rm A}/r_{\rm o}). 
\end{equation}

\noindent With measurements of $v_{\rm obs}$ and $r_{\rm obs}$,
together with an estimate of the outflow velocity $v_{\rm flow}$, from
equations 1 and 2 one can estimate the Alfv\'en radius,
$r_{\rm A}$, i.e.:

\begin{equation}
  r_{\rm A} \sim 2^{1/2} v_{\rm obs} r_{\rm obs} / v_{\rm flow}.
\end{equation}

\noindent Disk wind models typically predict a value of 2-5 for the
ratio $r_{\rm A}/r_{\rm o}$.  If we assume a value of 3, then an
estimate of $r_{\rm A}$ yields an estimate for the footprint radius,
$r_{\rm o}$.

Finally, from Kepler's law ($v_{\rm o} = (G M / r_{\rm o})^{1/2}$), we
may estimate the mass of the central star:

\begin{equation}
 G M \sim 2^{-1/2} v_{\rm obs} r_{\rm obs} v_{\rm flow} (r_{\rm o}/r_{\rm A})^3.
\end{equation}

\noindent In this way the footprint radius for the disk wind and the
mass of the central object may be derived using observable quatities,
$v_{\rm obs}$, $r_{\rm obs}$ and $v_{\rm flow}$, provided the theory
is correct.  For \irasa\ our slit spectra imply a toroidal velocity
$v_{\rm obs} \sim 5$~\kms\ at a radius $r_{\rm obs} \sim 3
\times 10^{14}$~m. The poloidal velocity of the molecular flow traced
in H$_2$ is difficult to measure because the bipolar nebula lies close
to the plane of the sky.  We assume a canonical speed of
100~\kms\ which, judging from our observations of other PPN, does
not seem unreasonable. These values lead to an Alfv\'en radius of $\sim
134$~AU and a source mass of 30~\Msolar .

Both seem to be somewhat on the high side (e.g. Frank \& Blackmann
2004; Kwok 1993).  Because we are observing the molecular component of
the flow, a lower flow velocity might be considered more approriate.
This would serve to reduce $M$, though $r_{\rm A}$ (and $r_{\rm o}$)
would then increase.  Large Alfv\'en and footprint radii might be
expected for the molecular component of a disk wind, since molecular
hydrogen would probably not survive acceleration close to the central
object.  Alternatively, $r_{\rm obs}$ could be overestimated,
particularly given the modest spatial resolution of our observations,
and the fact that the bow shock will be broader than the underlying
outflow.  $M$ and $r_{\rm A}$ are both directly proportional to the
observed flow radius; reducing $r_{\rm obs}$ would lower both
quantities.

The interaction of the fast wind with the AGB envelope could of course
produce asymmetries across the width of the flow.  In particular,
entrainment along one side of the flow could decelerate gas
along the flow edge producing skewed velocities that mimic rotation.
Such an effect would not be expected in both lobes of the flow,
however, and probably would not explain the gradual increase in
velocity seen across the {\em three} slit positions.

We conclude that, give the limited spatial resolution afforded by our
data, \irasa\ is certainly a candidate for nebula rotation, one that
deserves further study in a number of emission lines with a
high-resolution adaptive-optics spectrometer.

%

\section{Conclusions} 
 
From our high spectral-resolution near-IR spectroscopy we arrive at
the following conclusions:

\begin{enumerate}

\item{In bipolar PPN, H$_2$ emission is excited in -- and largely
confined to -- molecular shocks in the bipolar lobes.  The emission is
probably excited in shocks between the slow-moving AGB ejecta and the
much faster, post-AGB wind, specifically in the dense, swept-up layer
behind the forward shock at the fast wind/AGB wind interface (Frank
1994).  H$_2$ velocities are certainly more comparable to the velocity
of the detached AGB envelope than the fast wind.   }

\item{Unlike Garc\'{\i}a-Hern\'andez et al (2002), we find (albeit in
a very limited sample biased toward {\em bipolar} PPN) bright H$_2$
emission in PPN associated with early and reasonably late-type central
stars.  Our sample includes one O, two B and two G-type objects.
However, we did fail to detect H$_2$ along the axis of a bipolar
PPN associated with an M-type star.}

\item{H$_2$ emission is associated with PPN, regardless of whether
they exhibit \brg\ in emission or absorption.  This is, at least in
part, due to the fact that the H$_2$ and HI regions are physically
separate.}

\item{Our H$_2$ observations of \irasd , and both H$_2$ and [FeII]
observations of M~1-92, are convincingly modeled with molecular bow
shocks viewed roughly side-on.  Line widths, peak velocities, and
double-peaked profiles are reproduced.  The models also predict a
spatial offset between the H$_2$ and [FeII] emission regions, the
former being excited in the oblique bow flanks, the latter at the bow
apex. There is evidence for such an offset in M~1-92, although
additional observations at higher spatial resolution in these and other
PPN are warranted.}

\item{We find possible evidence for outflow rotation in one target,
\irasa .  Rotation is expected if the flow is driven
magneto-centrifugally from the surface of a disk.  Spectroscopy at high
spatial and spectral resolution would be extremely useful, with a
system that operates in the near-IR (such as {\em NACO} at the {\em
VLT}) to confirm our results and to search for similar trends in other
targets, since rotation would validate disk wind models, which have
recently been invoked to explain other aspects of outflows from
evolved stars.}

\end{enumerate}

\section{Acknowledgments} 
 
We thank the {\em Telescope System Specialists} at UKIRT, the numerous
observers who obtained data for this project through flexible
scheduling, and the referee for his/her thorough review of this
article.  The UKIRT is operated by the Joint Astronomy Centre on
behalf of the U.K. Particle Physics and Astronomy Research Council.
This research has also made use of the SIMBAD database, operated at
CDS, Strasbourg, France, and the NASA ADS database hosted by the
Harvard-Smithsonian Center for Astrophysics.

\label{lastpage}
\end{document}